\documentclass[aps,reprint,superscriptaddress,amsmath,amssymb]{revtex4-2}

\usepackage{graphicx}
\usepackage[colorlinks, citecolor=red]{hyperref}
\usepackage{siunitx}
\usepackage{braket}
\usepackage{multirow}
\usepackage{float}
\usepackage{epstopdf}
\usepackage{lineno}

\begin{document}


\title{Study of $s$- and $d$-wave intruder strengths in $^{13}{\rm B}_{\rm g.s.}$ via a $p(^{13}{\rm B},d)^{12}{\rm B}$ reaction}


\author{W. Liu}
\author{J. L. Lou}
\email[]{jllou@pku.edu.cn}
\author{Y. L. Ye}
\author{Z. H. Li}
\author{Q. T. Li}
\author{H. Hua}
\author{X. F. Yang}
\author{J. Y. Xu}
\affiliation{School of Physics and State Key Laboratory of Nuclear Physics and Technology, Peking University, Beijing 100871, China}
\author{H. J. Ong}
\affiliation{Research Centre for Nuclear Physics, Osaka University, Ibaraki, Osaka 567-0047, Japan}
\affiliation{CAS Key Laboratory of High Precision Nuclear Spectroscopy, Institute of Modern Physics, Chinese Academy of Sciences, Lanzhou 730000, China}
\affiliation{School of Nuclear Science and Technology, University of Chinese Academy of Sciences, Beijing 100080, China}
\author{D. T. Tran}
\affiliation{Research Centre for Nuclear Physics, Osaka University, Ibaraki, Osaka 567-0047, Japan}
\affiliation{Institute of Physics, Vietnam Academy of Science and Technology, Hanoi 10000, Vietnam}
\author{N. Aoi}
\author{E. Ideguchi}
\affiliation{Research Centre for Nuclear Physics, Osaka University, Ibaraki, Osaka 567-0047, Japan}
\author{D. Y. Pang}
\affiliation{School of Physics, Beijing Key Laboratory of Advanced Nuclear Materials and Physics, Beihang University, Beijing 100191, China}
\author{C. X. Yuan}
\affiliation{Sino-French Institute of Nuclear Engineering and Technology, Sun Yat-Sen University, Zhuhai 519082, China}
\author{S. M. Wang}
\author{Y. Jiang}
\author{B. Yang}
\author{Y. Liu}
\author{J. G. Li}
\author{Z. Q. Chen}
\author{J. X. Han}
\author{S. W. Bai}
\author{G. Li}
\author{K. Ma}
\author{Z. W. Tan}
\author{H. Y. Zhu}
\author{B. L. Xia}
\affiliation{School of Physics and State Key Laboratory of Nuclear Physics and Technology, Peking University, Beijing 100871, China}


\date{\today}

\begin{abstract}
Experimental results of the $p(^{13}{\rm B},d)^{12}{\rm B}$ transfer reaction to the low-lying states in $^{12}$B are reported.
The optical potential parameters for the entrance channel are extracted from the elastic scattering $p$($^{13}{\rm B}$, $p$) measured in the same experiment, while those for the exit channel are global ones.
Spectroscopic factors associated with the $p$-, $s$-, and $d$-wave neutron transfer to the known $^{12}$B states, are extracted by comparing the deuteron angular distributions with the calculation results.
The separated $s$- and $d$-wave intruder strengths in $^{13}{\rm B}_{\rm g.s.}$ were determined to be $10(2)\%$ and $6(1)\%$, respectively, which follow roughly the systematics for the $N$ = 8 neutron-rich isotones.
The measured total intruder strength is in good agreement with the shell model calculation, while the individual ones evolve quite differently.
Particularly, the sudden change of the $d$-wave intensity between $^{13}$B and $^{12}$Be needs further theoretical interpretation.
\end{abstract}


\maketitle


\section{\label{sec:introduction}Introduction}

Isospin dependence of nuclear structure and shell evolution is an interesting question of the finite fermi system, which is composed of certain numbers of protons and neutrons \cite{Otsuka2020}.
For light stable nuclei with small $N/Z$ asymmetry, the existence of the magic numbers can be well explained by the traditional shell model \cite{Mayer,Haxel}.
However, with increasing $N/Z$ ratios, unusual rearrangements of single-particle orbitals emerge in the exotic nuclei far from the $\beta$-stability line \cite{Liu2020, Chen2018, Chen20182, Tanihata2008, Tanaka2017, Kanungo2015,Hoffman2014,Hoffman2016}.

For the ground state (g.s.) of stable $N$ = 8 nuclei, the $0p0h$ configuration of ${(1p_{1/2})}^2_\nu$ is predominant according to the conventional shell model. However, the intruder $2p2h$ configurations of ${(p_{1/2})}^{-2}_{\nu}{(2s_{1/2})}^2_\nu$ ($s$-wave intruder) and ${(p_{1/2})}^{-2}_{\nu}{(1d_{5/2})}^2_\nu$ ($d$-wave intruder) have been widely investigated and reported for the neutron-rich $N$ = 8 nuclei \cite{Chen2018, Chen20182, Tanihata2008, Tanaka2017, Kanungo2015, Iwasaki2009, Tran2018, Ota2008}. These kinds of intrusion can lead to the breakdown of the magic number $N$ = 8.
For the Borromean nucleus $^{11}$Li, large $s$-wave intruder components in its g.s. ($\SI{45}{\percent}$) \cite{Tanihata2008} as well as in its low-lying excited states were observed \cite{Tanaka2017, Kanungo2015},
and theoretically attributed to the effect of the tensor and pairing forces \cite{Myo2007}.
In the nucleus with one more proton, namely $^{12}$Be, the $s$-wave component decreases to \SI{19}{\percent}, whereas the $d$-wave intrusion was found to be dominant \cite{Chen2018}.
The sudden increase of $d$-wave component in $^{12}{\rm Be}_{\rm g.s.}$ may be associated with the pairing interaction and deformation of the core \cite{Chen20182}.
For the nucleus with even two more protons, namely $^{14}$C, the g.s. is composed of only $\SI{1.3}{\percent}$ $s$-wave and slightly higher $d$-wave ($\SI{8.4}{\percent}$) components, as obtained in a $(p,d)$ transfer reaction \cite{Cecil1975}, indicating the restoration of the $N$ = $8$ magic number.
As a member of the $N$ = $8$ isotonic chain, between $^{12}$Be and $^{14}$C, $^{13}$B offers an intriguing opportunity to systematically understand the neutron shell evolution as a function of proton number $Z$.

If we neglect the contributions from $1d_{3/2}$ and higher orbitals, and keep $^{11}$B as the core, the wave function of $^{13}$B$\rm_{g.s.}$ can be written as
$\ket{\rm g.s.}=\nu[a(1p_{1/2})^2+b(2s_{1/2})^2+c(1d_{5/2})^2]$, where $a$, $b$, and $c$ stand for the spectroscopic amplitudes of $p$-, $s$-, and $d$-wave, respectively.
$p$-wave was found  to be dominant in $^{13}{\rm B}_{\rm g.s.}$ based on the measurement of its magnetic dipole moment \cite{Williams1971} and large $p$-wave spectroscopic factors (SFs) obtained from knockout \cite{Sauvan2004} and transfer \cite{Lee2010} reactions.
A \SI{33}{\percent} $s$-wave intruder strength in $^{13}{\rm B}_{\rm g.s.}$ was deduced from a $^{14}$Be $\beta$-decay experiment, in which the emissions of $\beta$-delayed neutrons with very low energies from the $1^+_1$ state were successfully measured \cite{Aoi2002}.
In a charge-exchange reaction experiment of $^{13}{\rm C}(t,^{3}{\rm He})^{13}{\rm B}$ \cite{Guess2009}, the wave function of $^{13}$B$\rm_{g.s.}$ was determined to be $\ket{^{13}{\rm B}}_{\rm g.s.}=0.871\ket{0\hbar\omega}+0.491\ket{2\hbar\omega}$,
where $0\hbar\omega$ and $2\hbar\omega$ correspond to the normal $p$-wave and the intruder $sd$-wave components, respectively.  This wave function gives a $24$\% intruder strength (a sum of $s$- and $d$-wave strengths), which is obviously smaller than the single $s$-wave intensity of $33$\% from the $\beta$-decay experiment.
Therefore, it is necessary to further investigate the $s$- and $d$-wave intruder intensities in $^{13}{\rm B}_{\rm g.s.}$ .

Single-neutron transfer reaction, which can selectively populate the states of interest,
is a powerful probe to investigate the single-particle strengths in a target nucleus \cite{Liu2020}.
In present work,  $p(^{13}{\rm B},d)^{12}{\rm B}$ transfer reaction is adopted to populate the well-known low-lying states in $^{12}{\rm B}$.
When a $p$-wave neutron is picked up by protons, the g.s. ($1^+$) and first excited state ($2^+$, \SI{0.953}{\mega\electronvolt}) in $^{12}$B will be populated.
The ratio of SFs for these two states should be similar to that obtained in the one-neutron knockout reaction \cite{Sauvan2004}.
If a neutron in the $2s_{1/2}$ orbital is transferred, the coupling of a residual $2s_{1/2}$ neutron with a $1p_{3/2}$ proton would lead to $(2,1)^-$ doublet at \SI{1.674}{} and \SI{2.621} {\mega\electronvolt}.
With a $1d_{5/2}$ neutron transferred, the configuration of $(1p_{3/2})^1_\pi\otimes(1d_{5/2})^1_\nu$ gives the $3^-_1$ (\SI{3.389} {\mega\electronvolt}), $1^-_2$ (\SI{4.302} {\mega\electronvolt}), $2^-_2$ (\SI{4.406} {\mega\electronvolt}), and $4^-_1$ (\SI{4.523} {\mega\electronvolt}) states.
The single-particle properties of the states in $^{12}$B mentioned above have been studied through several different $d(^{11}{\rm B},p)^{12}{\rm B}$ reactions \cite{Back2010, Lee2010, Belyaeva2018}.
Therefore, the population of negative-parity states in $^{12}$B via the $p(^{13}{\rm B},d)$ reaction will provide direct evidence for the $s$- and $d$-wave intrusions in ${\rm^{13}B_{g.s.}}$, with the intruder strengths determined directly from the corresponding SFs.

In this article, the experimental results of the first $p(^{13}{\rm B},d)^{12}{\rm B}$ transfer reaction using a radioactive beam of $^{13}$B at 23 MeV/nucleon are presented.
The experimental setup and results for reactions on the polyethylene target are given in Sec.~\ref{sec:experimentaldetails} and Sec.~\ref{sec:results}, respectively.
A brief summary is given in Sec.~\ref{sec:summary}.

\section{\label{sec:experimentaldetails}Experimental Details}

The experiment was performed at the EN-course beam line at Research Center for Nuclear Physics (RCNP), Osaka University \cite{Shimoda1992, Ong2014}.
A 23-MeV/nucleon $^{13}$B secondary beam was produced from a 58-MeV/nucleon $^{18}$O primary beam impinging on a 3.8-mm-thick $^{9}$Be target.
The beam was purified by the electromagnetic separator after punching through a 3.07-\SI{}{\milli\meter}-thick aluminium degrader. The beam purity was about \SI{98}{\percent}, with approximately ${2.0 \times 10^{4}}$ particles per second.
The momentum spread was limited, by a slit, down to ${\Delta p/p\leq\SI{1.5}{\percent}}$ to reduce the energy
dispersion of the secondary beam.

\begin{figure}
\includegraphics[width=1.0\hsize]{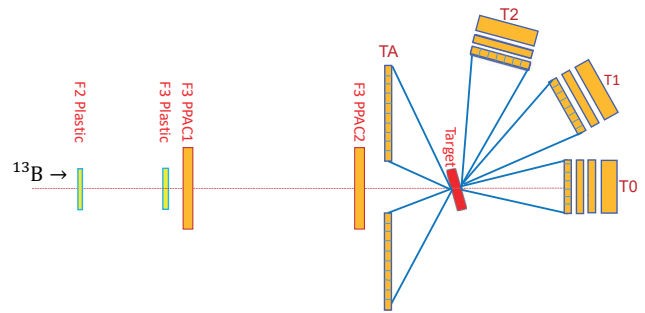}
\caption{\label{fig:experimentalSetup} Schematic view of the experimental setup.}
\end{figure}

The experimental setup is schematically shown in Fig.~\ref{fig:experimentalSetup}.
The secondary beam was identified by the time-of-flight (TOF), which was provided by two plastic scintillation detectors (F2 Plastic and F3 Plastic), and energy losses in F3 Plastic.
Two x-y position-sensitive parallel plate avalanche counts (PPACs) were employed to track the $^{13}$B beam before the physical target. The position resolution of two PPACs was about \SI{1.5} {\milli\meter} (FWHM), and the distance between them was about \SI{772} {\milli\meter}, resulting in an angular resolution within \SI{0.28}{\degree}.
A \SI{6.76} mg/cm$^2$ polyethylene target ${\rm (CH_2)}_n$ and a \SI{3.98} mg/cm$^2$ deuterated polyethylene target ${\rm (CD_2)}_n$ were installed.
Both targets were rotated \SI{20}{\degree} with respect to the beam direction in order to reduce the energy losses in the target of the low-energy charged particles emitting to the telescopes T2 and T1 (as shown in Fig.~\ref{fig:experimentalSetup}).

Four sets of charged particle telescopes (T0, T1, T2, TA) were employed in the large scattering chamber, as shown in Fig.~\ref{fig:experimentalSetup}.
T0, T1, and T2 were used to detect the residual nuclei around the beam direction, the reaction-produced light particles, and the elastically scattered protons and deuterons at large angles, respectively.
TA was placed at backward angle to detect the protons from the $d(^{13}{\rm B},p)^{14}{\rm B}$ reaction.

\begin{figure}
\includegraphics[width=1.0\hsize]{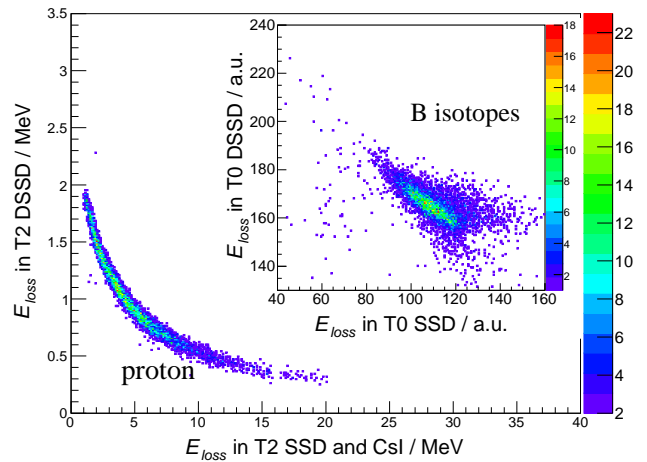}
\caption{\label{fig:particleIdentificationT2} Protons elastically scattered from $^{13}$B in the particle identification (PID) spectrum measured by T2. The inset figure shows the PID spectrum measured by T0, where the Boron isotopes are shown in coincidence with protons.
.}
\end{figure}

T0, consisting of a double-sided silicon strip (DSSD) detector with a thickness of \SI{1000} {\micro\meter}, two large size silicon detectors (SSDs), and a layer of 4-cm-thick CsI(Tl) crystals read out by photodiodes, was centered at \SI{0}{\degree} with respect to the beam direction.
T2, which was composed of a 60-mm-thick DSSD and a SSD, was installed around \SI{70}{\degree} with respect to the beam direction.
An array of 4-cm-thick CsI(Tl) crystals was placed behind SSD to stop high energy particles.
T1 was placed around \SI{31}{\degree}, with the same composition as T2.
Each DSSD used in these three sets of telescopes is divided into 32 strips on both sides and has an active area of ${64 \times 64}$ \SI{}{\milli\meter\squared}.
Each SSD has the same active size as the DSSDs while its nominal thickness is \SI{1500} {\micro\meter}.
TA is a set of the annular double-sided silicon strip detector (ADSSD) composed of six sectors. The distances between the center of target and the first layer of each telescope are \SI{200}{}, \SI{150}{}, \SI{150}{}, and \SI{190} {\milli\meter} for T0, T1, T2, and TA, respectively.

The angular resolution of the telescopes T2 (and T1) approximates \SI{0.9}{\degree} (FWHM), taking into consideration the angular resolution resulting from the PPACs and strip width of DSSD.
The energy resolution of the silicon detectors was less than \SI{1}{\percent} for $\alpha$ particles at \SI{5.486} {\mega\electronvolt}, which was sufficient to discriminate the nuclei lighter than carbon, with the standard $\Delta E-E$ method.

In this paper, we will focus on the $(p,p)$ elastic scattering and $(p,d)$ transfer reaction between $^{13}$B and the ${\rm (CH_2)}_n$ target.
The light particles detected in T2 or T1 are analysed in coincidence with the residual nuclei detected in T0.
Data from the coincidence of T0 and TA are not involved in the following sections.

\section{\label{sec:results}Results}

\subsection{\label{subsec:particleidentification}Particle Identification}

\begin{figure}
\includegraphics[width=1.0\hsize]{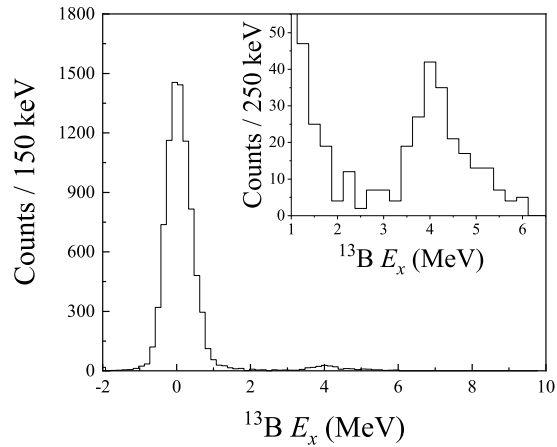}
\caption{\label{fig:elasticScatteringErelCH} Excitation energy spectrum of $^{13}$B reconstructed from the energy and momentum of the scattered protons measured by T2. The inset figure shows the energy spectrum for inelastic scattering.}
\end{figure}

The experiment was performed in inverse kinematics, and different reaction channels were discriminated by the coincidence detection of the boron isotopes in T0 and light particles in other telescopes.
As shown in Fig.~\ref{fig:particleIdentificationT2}, protons detected in T2 are clearly seen.
They are mainly from the elastic scattering of $^{13}$B on the ${\rm (CH_2)}_n$ target.
Besides elastic scattering, some of the protons are from the inelastic scattering.
The first and second positive-parity excited states of $^{13}$B at \SI{3.48}{} and \SI{3.68}{\mega\electronvolt}, which are formed by the excitation of one neutron from the $1p_{1/2}$ orbital to the $2s_{1/2}$ and the $1d_{5/2}$ orbital \cite{Back2010}, respectively, can be populated by the inelastic scattering.
As shown in Fig.~\ref{fig:elasticScatteringErelCH}, in the excitation energy spectrum of $^{13}$B reconstructed for the $(p,p^\prime)$ inelastic scattering, a wide peak was observed around \SI{3}{}-\SI{5} {\mega\electronvolt}.
Due to the limited acceptance of T2, higher excited states in $^{13}$B were not observed for the ${\rm (CH_2)}_n$ target.
The inset of Fig.~\ref{fig:particleIdentificationT2} shows the particle identification (PID) spectrum detected in T0 after gated on the protons detected in T2, in which $^{13}$B is predominant.

\begin{figure}
\includegraphics[width=1.0\hsize]{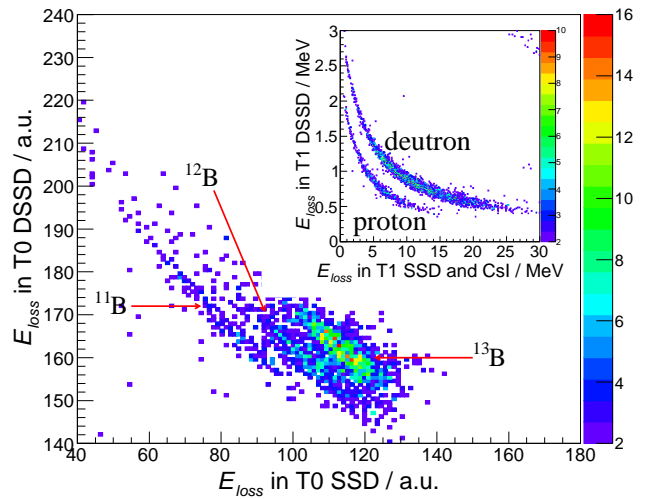}
\caption{\label{fig:particleIdentificationT1} Particle Identification spectrum of boron isotopes detected by T0 in coincidence with T1. The inset figure shows the PID measured by T1.}
\end{figure}

Fig.~\ref{fig:particleIdentificationT1} shows the boron isotopes detected in T0, with the gate of light particles measured in T1.
The hydrogen isotopes detected in T1 are shown in the inset of Fig.~\ref{fig:particleIdentificationT1}.
The neutron threshold of $^{12}$B is $S_n=\SI{3.370} {\mega\electronvolt}$.
For excited states above this threshold, the produced $^{12}$B will emit a neutron, and $^{11}$B was detected in T0.
Normally, $^{12}$B, and $^{11}$B will be seen and separated from $^{13}$B in T0, which is similar to the results in \cite{Jiang2018}.
$^{11}$B, $^{12}$B, and $^{13}$B are clearly seen, but are not separated well due to the poor energy resolution of the first layer SSD in T0.
The analysis of cross section in the following subsections was based on the coincidence of light charged particles in T1 or T2, and boron isotopes in Fig.~\ref{fig:particleIdentificationT1} or the inset of Fig.~\ref{fig:particleIdentificationT2}.

\subsection{\label{subsec:elasticscattering}Elastic Scattering}

\begin{figure}
\includegraphics[width=1.0\hsize]{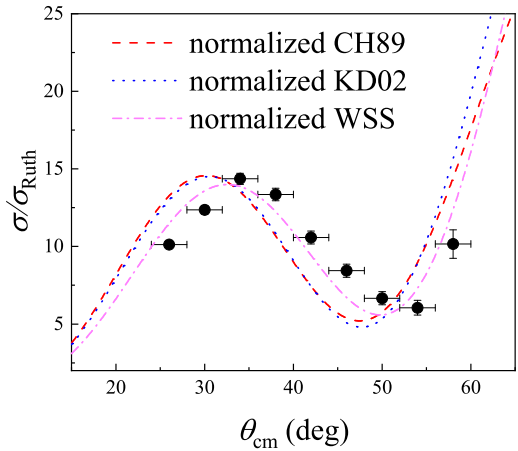}
\caption{\label{fig:elasticScatteringCrossScetionCH} Elastic scattering differential cross sections, relative to the Rutherford cross sections. Three sets of global optical potentials were employed for the theoretical calculations. See text for more details.}
\end{figure}

The elastic scattering differential cross section, as a ratio to the Rutherford cross section is shown in Fig.~\ref{fig:elasticScatteringCrossScetionCH}.
The count for each point, which corresponds to an angular range of \SI{2}{\degree} in the laboratory frame, is obtained by fitting the energy spectrum deduced from the energy and momentum of light particles detected in T2.
The peak center is zero. The resolution of 776 keV(FWHM) is in good agreement with the simulated result using the code GEANT4 \cite{Agostinelli2003}. Only the amplitude of the Gaussian function is left as a free parameter for each point.
The error bars are purely statistical.
The systematic error is less than \SI{9}{\percent}, considering the uncertainties in the geometrical efficiency determination, the thickness of the target, and the cuts on the PID spectra as shown in Fig.~\ref{fig:particleIdentificationT2}.

\begin{figure*}
\includegraphics[width=0.9\hsize]{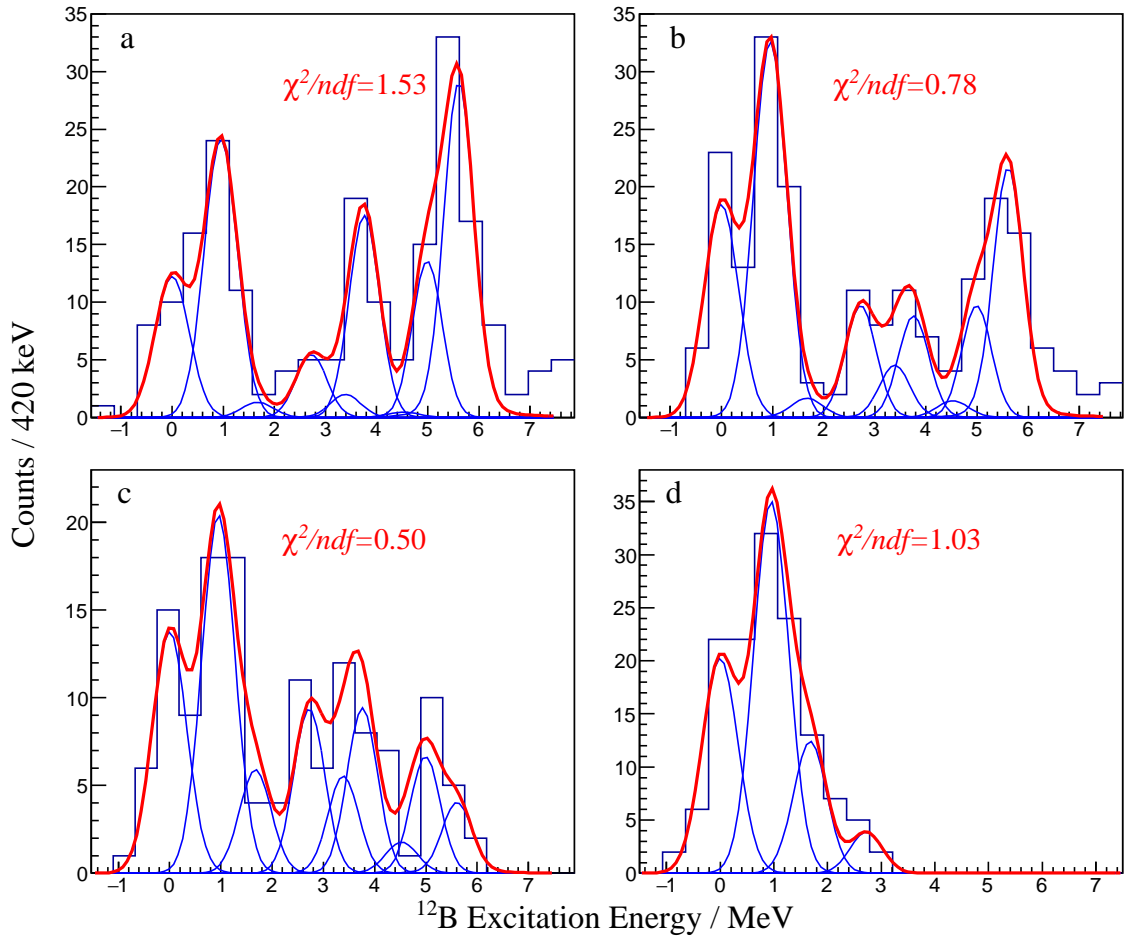}
\caption{\label{fig:excitationEnergySpectrumFor12BCS} Excitation energy spectra of $^{12}$B reconstructed from the energy and momentum of deuterons at (a) $23^\circ-27^\circ$, (b) $27^\circ-31^\circ$, (c) $31^\circ-35^\circ$, (d) $35^\circ-39^\circ$. The blue solid lines show the fitted spectrum for each state, and the red solid curves show the total fits, i.e. sums of blue solid curves. Note that the amplitudes for states above \SI{3}{\mega\electronvolt} are zero in (d). The number of degrees of freedom (ndf) corresponds to the number of points used in the fit minus the number of free parameters.}
\end{figure*}

Optical model is a powerful tool to describe the elastic scattering \cite{Dickhoff2019}.
Three sets of global optical potentials (OPs), CH89 \cite{Varner1991}, KD02 \cite{Koning2003}, and WSS \cite{Watson1969}, were applied in the calculations with the code FRESCO \cite{Thompson1988} for the proton angular distribution.
To better describe the experimental data, two normalization factors, $\lambda _{V}$ and $\lambda _{W}$, were introduced to the depths of the real ($V_V$) and the imaginary part ($W_V+W_S$), respectively \cite{Chen2016}.
The geometric parameters $r_0$ and $a$ were adopted the same as those of the global OPs.
The searching process for the renormalization factors was based on the $\chi^2$ minimization method. It shows that the normalized WSS potential describes the current data better.
The normalization factors and corresponding $\chi^2/n$ values are summarized in Table.~\ref{table:normalizationFactor}.
The uncertainties for $\lambda _{V}$ and $\lambda _{W}$ are given in the \SI{68}{\percent} confidence level with ${\chi}^2_{min}+2.3$ for the fitting procedure with two free parameters.

\begin{table} 
\caption{\label{table:normalizationFactor} Normalization factors applied to the real ($V_V$) and imaginary ($W_V+W_S$) central potentials of  CH89 \cite{Varner1991}, KD02 \cite{Koning2003}, and WSS \cite{Watson1969} for the $^{13}{\rm B}$ + $p$ elastic scattering. The uncertainties in these normalization factors are shown in the parentheses, corresponding to the standard deviation of the fit.}
\begin{ruledtabular}
\begin{tabular}{cccc}
global OP & $\lambda _{V}$ & $\lambda _{W}$ & $\chi^2/n$ \\
\colrule
CH89 & 1.17(1) & 1.07(2) & 62.59 \\
KD02 & 1.24(3) & 1.10(10) & 59.74 \\
WSS & 1.11(4) & 0.98(4) & 13.80 \\
\end{tabular}
\end{ruledtabular}
\end{table}

\subsection{\label{subsec:singleneutrontransfer} $p(^{13}{\rm B},d)^{12}{\rm B}$ Transfer Reaction}

The excitation energy spectrum of $^{12}$B was reconstructed from the energy and momentum of deuterons detected in T1 in coincidence with boron isotopes measured in T0.
As shown in Fig.~\ref{fig:excitationEnergySpectrumFor12BCS}, the spectra were fitted with nine known states of $^{12}$B, namely the g.s. and the excited states at $E_x$ = \SI{0.953}{}, \SI{1.674}{}, \SI{2.723}{}, \SI{3.389}{}, \SI{3.760}{}, \SI{4.523}{}, \SI{4.990}{}, and \SI{5.610} {\mega\electronvolt}.
The position for each peak was fixed.
The width of each peak is predominated by the detection resolution, which was simulated by using Geant4.
Only the amplitudes were left as free parameters.
All the bound states and unbound states locating above $S_n({\rm ^{12}B})=\SI{3.370} {\mega\electronvolt}$ were fitted with functions in Gaussian form, except for the \SI{5.610} {\mega\electronvolt} state.
Considering its large intrinsic width, a Breit–Wigner function convoluted with the simulated resolution of the detector system was adopted \cite{Tanaka2017} for this state.
In this experiment, both the energy and the time information of each detector were recorded.
Time cut was used when analysing the excitation energy spectra from the coincidence events.
The background contributions from reactions with carbon in the physical target of (CH$_2$)$_n$ were negligible, which were verified by the analysis with carbon and empty target.
Thus, no background was included in the fitting functions.

\begin{figure}
\includegraphics[width=0.99\hsize]{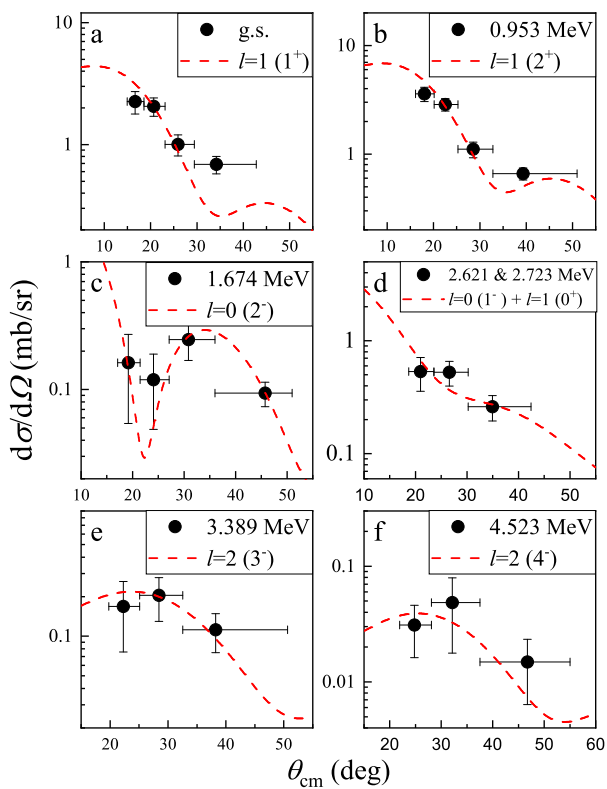}
\caption{\label{fig:b12CrossScetion} Differential cross sections of $p({}^{13}{\rm B},d)$ to the excited states at (a) 0.00, (b) 0.953, (c) 1.674, (d) 2.261 and 2.723, (e) \SI{3.389}{}, (f) \SI{4.523} {\mega\electronvolt}. The excitation energy, the spin-parity, and the transferred orbital angular momentum $l$, are given in the figure. The red dashed curves are the calculated differential cross sections for each state, multiplied by the corresponding SFs. The \SI{2.621} {\mega\electronvolt} state could not be separated from the \SI{2.723} {\mega\electronvolt} state in this experiment, so it was fitted as one peak in Fig.~\ref{fig:excitationEnergySpectrumFor12BCS} and decomposed by two components in (d). }

\end{figure}

Differential cross sections for the populated states of interest are shown in Fig.~\ref{fig:b12CrossScetion}.
The error bars are purely statistical.
The systematic error is less than \SI{11}{\percent}, taking into consideration the uncertainties in the geometrical efficiency determination, the thickness of the target, and the cuts of deuterons and borons on the PID spectra as shown in Fig.~\ref{fig:particleIdentificationT1}.

To extract the SFs, the distorted wave Born approximation calculation was performed with the code FRESCO.
Considering the relatively smaller $\chi^2$ value, the normalized WSS potential was adopted for the entrance channel. The systematic optical potential of Daehnick \textit{et al.} \cite{Daehnick1980} was used for the exit channel.
Reid soft-core potential \cite{Lassey1975} was employed to reproduce the binding energy of deuteron.
For the interaction between $^{12}$B and neutron, the Woods-Saxon potential was chosen.
The geometry parameters $r_0$ and $a$ were set to \SI{1.25}{\femto\meter} and \SI{0.65}{\femto\meter}, respectively.
The depths of binding potentials were adjusted to reproduce the binding energy of $^{13}$B.
As shown in Fig.~\ref{fig:b12CrossScetion}, the red dashed curves are the calculated differential cross sections for each state, which have been multiplied by the corresponding SFs extracted in this experiment.
Note that the triplet formed by the transfer of the $1d_{5/2}$ neutron, namely the \SI{4.302}{}, \SI{4.406}{}, and \SI{4.523} {\mega\electronvolt} states, are not resolved due to the limited $Q$-value resolution.
Considering the limited statistics of the differential cross sections for this triplet (as shown in Fig.~\ref{fig:b12CrossScetion}f), it is difficult to decompose it by using three components.
The SF used in Fig.~\ref{fig:b12CrossScetion}f was extracted by comparing the differential cross sections to the theoretical ones calculated for the \SI{4.523} {\mega\electronvolt} state, which is the smallest among the theoretical calculations for these three states.
Thus, the extracted SF is the upper limit for those three states.

Since the experimental SFs are sensitive to the choice of OPs and practical experimental conditions \cite{Chen2018, Liu2020}, the relative SFs are more meaningful.
According to the sum rule \cite{Macfarlane1960}, the sum of all the $p$-, $s$-, and $d$-wave SFs, corresponding to the $l$ = 1, 0, and 2 neutron transfer in the $p$($^{13}$B$_{g.s.}$,$d$) reaction, should be equal to 2.0.
Based on this principle, the experimental SFs for single-particle states formed by $p$-, $s$-, and $d$-wave neutron were normalized to obtain the relative SFs, and the results are summarized in Table. \ref{table:spectroscopicfactor}.
The uncertainties for the SFs correspond to a \SI{68}{\percent} confidence level with ${\chi}^2_{min}+1$ except for the \SI{2.621} {\mega\electronvolt} state. The SF uncertainty for the \SI{2.621} {\mega\electronvolt} state was deduced from ${\chi}^2_{min}+2.3$, because two parameters were used in the fit to the corresponding differential cross sections.
The ratio of SFs for the g.s. and first excited state is in agreement with that obtained from the single-neutron knockout reaction \cite{Sauvan2004}.

\begin{table*} 
\caption{\label{table:spectroscopicfactor} Excitation energies and SFs for the low-lying states in $^{12}$B. The relative SFs are extracted from the present $p$($^{13}$B,$d$) reaction to the lowing-lying states in $^{12}$B and the corresponding uncertainties are from the fit to the differential cross sections for each state based on the $\chi^2$ minimization method. Comparing with the experimental results, the shell model calculation results with the WBP \cite{Warburton1992} interaction and the latest YSOX interaction \cite{Yuan2012} are also listed. }
\begin{ruledtabular}
\begin{tabular}{cccccccc}
\multirow{2}{*}{spin-parity} & \multirow{2}{*}{orbital} & \multicolumn{2}{c}{Exp.} & \multicolumn{2}{c}{YSOX} & \multicolumn{2}{c}{WBP} \\ \cline{3-4} \cline{5-6} \cline{7-8}
 & & $E_x$ $(\SI{} {\mega\electronvolt})$ & $\rm{SF_{rel}}$ &
 $E_x$ $(\SI{}{\mega\electronvolt})$ & $\rm{SF}$ & $E_x$ $(\SI{}{\mega\electronvolt})$ & $\rm{SF}$ \\
\colrule
$1^+_1$ & $1p_{1/2}$ & 0.000 & $0.55(5)$ & 0.000 & 0.49 & 0.000 & 0.53 \\
$2^+_1$ & $1p_{1/2}$ & 0.953 & $1.12(8)$ & 1.395 & 0.96 & 1.631 & 1.04 \\
$2^-_1$ & $2s_{1/2}$ & 1.674 & $0.12(2)$ & 1.490 & 0.04 & 2.885 & 0.003 \\
$1^-_1$ & $2s_{1/2}$ & 2.621 & $0.09(4)$ & 2.222 & 0.02 & 3.702 & 0.003 \\
$3^-_1$ & $1d_{5/2}$ & 3.389 & $0.10(2)$ & 2.842 & 0.10 & 4.193 & 0.03 \\
$1^-_2$ & $1d_{5/2}$ & 4.302 &  & 3.902 & 0.002 & 4.102 & 0.001 \\
$2^-_2$ & $1d_{5/2}$ & 4.460 &  & 3.359 & 0.03 & 4.362 & 0.006 \\
$4^-_1$ & $1d_{5/2}$ & 4.523 & $0.02(1)$ \footnote{Upper limit for the sum of SFs of \SI{4.302}{}, \SI{4.460}{}, and \SI{4.523}{\mega\electronvolt} states, see text for more details.} & 3.889 & 0.06 & 4.348 & 0.02 \\
\end{tabular}
\end{ruledtabular}
\end{table*}

\subsection{\label{subsec:configuration} $s$- and $d$-wave strengths in $^{13}{\rm B}_{\rm g.s.}$}
By ignoring the contributions from $1d_{3/2}$ and higher orbitals, the g.s. of $^{13}$B could be written as $\ket{\rm g.s.}=\nu[a(1p_{1/2})^2+b(2s_{1/2})^2+c(1d_{5/2})^2]$, with the normalization relation of $a^2+b^2+c^2=\alpha+\beta+\gamma=1$, where $\alpha$, $\beta$, and $\gamma$ are the $p$-, $s$-, and $d$-wave intensities, respectively.
The ratio of $\alpha$, $\beta$, and $\gamma$ is deduced from the sum of the relative SFs of low-lying states in $^{12}$B populated by the transfer of the $p$-, $s$-, and $d$-wave neutron in the present $p$($^{13}$B,$d$) reaction.
Based on the studies of $^{14}{\rm C}_{\rm g.s.}$ via $(p,d)$ transfer reaction \cite{Cecil1975, Yasue1990}, contributions of $p$-, $s$-, and $d$-wave strengths from high-lying excited states are small and can be ignored.
Combining the ratio of $\alpha$, $\beta$ with $\gamma$ and $\alpha+\beta+\gamma=1$, we obtained $\alpha= 84(7)\%$, $\beta=10(2)\%$, and $\gamma=6(1)\%$.
The intruder $s$-wave strength in $^{13}{\rm B}_{\rm g.s.}$ is nearly twice that of the $d$-wave, which may imply indirectly the inversion of $2s_{1/2}$ and $1d_{5/2}$ orbitals, comparing with the conventional shell model.
The intruder $s$-wave intensity of 10(2)\% extracted directly from our experiment is dramatically lower than 33\%, which was indirectly deduced from the $^{14}$Be $\beta$-decay experiment \cite{Aoi2002}. However, the sum of intruder $s$- and $d$-wave strength is 16(3)\%, which is close to 24\% obtained from the charge-exchange reaction ~\cite{Guess2009}.

\begin{figure}
\includegraphics[width=0.99\hsize]{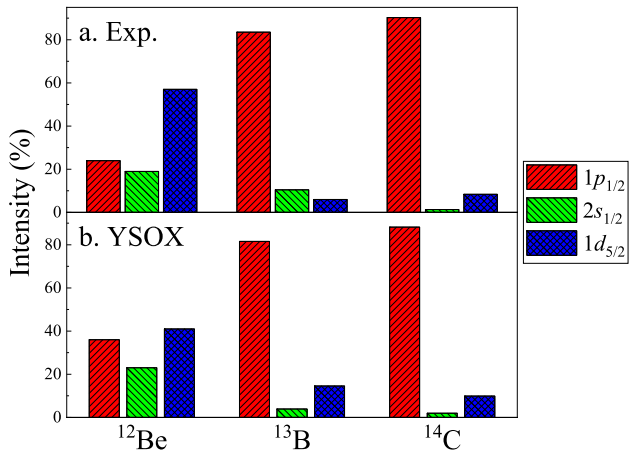}
\caption{\label{fig:intensity} (a)Individual $p$-, $s$-, and $d$-wave intensities in the ground state of $^{12}$Be \cite{Chen2018}, $^{13}$B (this work) and $^{14}$C \cite{Cecil1975}. (b) Shell model calculations with YSOX interaction in full $p$-$sd$ model space. The results for $^{12}$Be are from Ref. \cite{Chen2018}.}
\end{figure}

Fig.~\ref{fig:intensity}a shows the experimental results of the $p$-, $s$-, and $d$-wave intensities in $^{12}$Be$\rm_{g.s.}$ \cite{Chen2018}, $^{13}$B$\rm_{g.s.}$ and $^{14}$C$\rm_{g.s.}$ \cite{Cecil1975}.
It can be seen that the results of $^{13}$B$\rm_{g.s.}$ are consistent with the systematical trends of the $N$ = 8 isotones.
The intensity of the $p$-wave increases with increasing $Z$ value, whereas the intensities of the intruder $s$- and $d$-wave decrease dramatically.
With difference of merely one proton, the $d$-wave intruder strength changes suddenly from $^{12}$Be to $^{13}$B, but remains nearly a constant from $^{13}$B to $^{14}$C.
The sudden change of the intruder configuration between $^{12}$Be and $^{13}$B is interesting and needs further theoretical interpretation.

In comparison with the experimental results, the $p$-, $s$-, and $d$-wave intensities calculated from the shell model with the YSOX interaction \cite{Yuan2012} are shown in Fig.~\ref{fig:intensity}b.
The calculations were performed in a full $p$-$sd$ model space.
The measured $p$-wave intensities (red bars) for these nuclei can be fairly reproduced by the calculations.
However, the measured $d$-wave intensity (blue bar) in $^{12}$Be evolves quite differently from the shell model calculations.
The $s$-wave intensities (green bars) in $^{12}$Be and $^{13}$B are also inconsistent with the calculated results.
The total $sd$-wave intruder strength of 16(3)\% from this experiment is in good agreement with 15\% from the shell model calculation.
This is an indication of the necessity to measure individual components of the mixing configuration in order to further advance the theoretical approaches.

We also performed calculations within the Gamow coupled-channel (GCC) approach by assuming $^{13}$B as a deformed $^{11}$B core plus two neutrons, together with coupling to the continuum.
The calculation adopted the same interaction and model space as in Ref.~\cite{Wang2019}, which has successfully reproduced the low-lying excited states in $^{12}$Be.
As a result, the calculated intruder $s$- and $d$-wave strengths in $^{13}{\rm B}_{\rm g.s.}$ are only 2.06$\%$ and 2.35$\%$, respectively, by treating the valence proton as a speculator.
If we considered the effect of $pn$ interaction, the total intruder strength would even decrease to less than 2$\%$. For $^{12}{\rm Be}_{\rm g.s.}$, the calculated $d$-wave intensity of 25$\%$ is only half of the experimental data, while the $s$-wave strength of 20$\%$ is nearly the same as the measured value ~\cite{Wang2019}.
This implies different structures in $^{13}$B and $^{12}$Be, particularly the restoration of $N$ = 8 magic shell. In order to better reproduce the experimental results, the strength of nucleon-nucleon interaction could vary in different nuclear mediums.

\section{\label{sec:summary}Summary}

In summary, the $p(^{13}{\rm B},d)^{12}{\rm B}$ transfer reaction was performed with a 23-MeV/nucleon $^{13}$B beam.
With the optical potential extracted from the elastic scattering data measured in the same experiment, the relative SFs associated with the $^{13}{\rm B}_{\rm g.s.}$ configuration were determined through the measurements of a single-neutron transfer reaction to the known $^{12}$B states.
The $s$- and $d$-wave intruder strengths in $^{13}$B$\rm_{g.s.}$ are deduced quantitatively, which are consistent with the systematics of $N$ = 8 neutron-rich isotones.
The total intruder $2p2h$ strength including both $s$- and $d$-wave is in good agreement with the shell model calculations.
However, the measured individual ones evolve quite differently compared with the calculations.
Particularly, the sudden increase of the $d$-wave intensity moving from $^{13}{\rm B}_{\rm g.s.}$ to $^{12}{\rm Be}_{\rm g.s.}$ needs further theoretical interpretation.
The present work demonstrates that the single-nucleon transfer reaction in inverse kinematics is a sensitive tool to investigate the configuration mixing, including the intruder phenomena, in unstable neutron-rich nuclei.

\begin{acknowledgments}

We gratefully thank the RCNP accelerator group for providing $^{18}$O primary beam and the RCNP technical staff for assistance.
This work was supported by the National Key R$\&$D Program of China (Grant No. 2018YFA0404403), and the National Natural Science Foundation of China (Contract Nos. 11775004, U1867214, 11775013, 11775316, 11875074, 11961141003, and U2067205).

\end{acknowledgments}

\bibliography{ref}

\end{document}